\documentstyle[buckow]{article}

\makeatletter \@addtoreset{equation}{section} \makeatother

\def\bfone{\relax{\rm 1\kern-.35em 1}}
\def\bfzero{\relax{\rm I\kern-.18em 0}}
\def\inbar{\vrule height1.5ex width.4pt depth0pt}
\def\IC{\relax\,\hbox{$\inbar\kern-.3em{\rm C}$}}
\def\ID{\relax{\rm I\kern-.18em D}}
\def\IF{\relax{\rm I\kern-.18em F}}
\def\IK{\relax{\rm I\kern-.18em K}}
\def\IH{\relax{\rm I\kern-.18em H}}
\def\II{\relax{\rm I\kern-.17em I}}
\def\IN{\relax{\rm I\kern-.18em N}}
\def\IP{\relax{\rm I\kern-.18em P}}
\def\IQ{\relax\,\hbox{$\inbar\kern-.3em{\rm Q}$}}
\def\IR{\relax{\rm I\kern-.18em R}}
\def\IG{\relax\,\hbox{$\inbar\kern-.3em{\rm G}$}}
\font\cmss=cmss10 \font\cmsss=cmss10 at 7pt
\def\ZZ{\relax\ifmmode\mathchoice
{\hbox{\cmss Z\kern-.4em Z}}{\hbox{\cmss Z\kern-.4em Z}}
{\lower.9pt\hbox{\cmsss Z\kern-.4em Z}} {\lower1.2pt\hbox{\cmsss
Z\kern .4em Z}}\else{\cmss Z\kern-.4em Z}\fi}

 \def\cD{{\cal D}}
    \def\cF{{\cal F}} \def\cG{{\cal G}}

\def\Coe#1.#2.{{#1\over #2}}

\def\coe#1.#2.{\relax{\textstyle {#1 \over #2}}\displaystyle}

\def\notin{\hbox{{$\in$}\kern-.51em\hbox{/}}}

\def\IE{\relax{{\rm I\kern-.18em E}}}

\def\IGam{\relax{{\rm I}\kern-.18em \Gamma}}

\def\IA{\relax{\hbox{{\rm A}\kern-.82em {\rm A}}}}

\begin {document}
\def\titleline{
 $AdS_6/CFT_5$ correpondence for $F(4)$ Supergravity
\footnote{The work of R. D'Auria and S. Vaul\'a has been
supported by EEC under TMR contract ERBFMRX-CT96-0045, the work
of S. Ferrara has been supported by the EEC TMR programme
ERBFMRX-CT96-0045 (Laboratori Nazionali di Frascati, INFN) and by
DOE grant DE-FG03-91ER40662, Task C.}}


\def\authors{
Riccardo D'Auria\1ad, Sergio Ferrara\2ad and Silvia Vaul\'a\1ad
}
\def\addresses{ \1ad Dipartimento di Fisica, Politecnico di
Torino,\\
 Corso Duca degli Abruzzi 24, I-10129 Torino\\
and Istituto Nazionale di Fisica Nucleare (INFN) - Sezione di
Torino, Italy\\ \2ad CERN Theoretical Division, CH 1211 Geneva 23, Switzerland\\
and Istituto Nazionale di Fisica Nucleare, Laboratori Nazionali
di Frascati, Italy\\


}
\def\abstracttext{
$F(4)$  supergravity, the gauge theory of the exceptional
six-dimensional Anti-de Sitter superalgebra, is coupled  to an
arbitrary number of vector multiplets whose scalar components
parametrize the quaternionic manifold $SO(4,n)/SO(4)\times
SO(n)$. By gauging the compact subgroup $SU(2)_d \otimes \cG$,
where $SU(2)_d$ is the diagonal subgroup of $SO(4)\simeq
SU(2)_L\otimes SU(2)_R$ (the$R$-symmetry group of six-dimensional
Poincar\'e supergravity) and $\cG$ is a compact group such that
$dim\cG = n$. The potential admits an $AdS$ background for $g=3m$,
as the pure $F(4)$-supergravity. The boundary $F(4)$
superconformal fields are realized in terms of a singleton
superfield (hypermultiplet) in harmonic superspace with flag
manifold $SU(2)/U(1)=S^2$. We analize the spectrum of short
representations in terms of  superconformal primaries and predict
general features of the K-K specrum of massive type IIA
supergravity compactified on warped $AdS_6\otimes S^4$.
} \large \makefront
\section{Matter coupled F(4) Supergravity and its scalar potential}

Let us set up a suitable framework for the discussion of the
matter coupled $F(4)$ supergravity theory and its gauging.
\noindent This will allow us to set up the formalism for the
matter coupling in the next section. Actually we will just give
the essential definitions of the Bianchi identities approach in
superspace , while all the relevant results, specifically the
supersymmetry transformation laws of the fields, will be given in
the ordinary space-time formalism.

First of all it is useful to discuss the main results of ref.
\cite{rom} by a careful study in superspace of the Poincar\'e and
AdS supersymmetric vacua.
 Let us recall the content of $D=6$, $N=(1,1)$ supergravity
multiplet:
\begin{equation}
(V^{a}_{\mu}, A^{\alpha}_{\mu}, B_{\mu\nu},\,\ \psi^{A}_{\mu},\,\
\psi^{\dot{A}}_{\mu}, \chi^{A}, \chi^{\dot{A}}, e^{\sigma})
\end{equation}
\noindent where $V^a_{\mu}$ is the six dimensional vielbein,
$\psi^{A}_{\mu},\,\ \psi^{\dot{A}}_{\mu}$ are left-handed and
right-handed four- component gravitino fields respectively, $A$
and $\dot{A}$ transforming under the two factors of the
$R$-symmetry group $O(4)\simeq SU(2)_L\otimes SU(2)_R$,
$B_{\mu\nu}$ is a 2-form, $A^{\alpha}_{\mu}$ ($\alpha=0,1,2,3$),
are  vector fields, $\chi^{A}, \chi^{\dot{A}}$ are left-handed and
right-handed spin $\frac{1}{2}$ four components dilatinos, and
$e^{\sigma}$ denotes the dilaton.\\ Our notations are as follows:
$a,b,\dots=0,1,2,3,4,5$ are Lorentz flat indices in $D=6$
$\mu,\nu,\dots=0,1,2,3,4,5$ are the corresponding world indices,
  $A,\dot{A}=1,2$. Moreover our metric is
$(+,-,-,-,-,-)$.\\ We recall that the description of the spinors
of the multiplet in terms of left-handed and right-handed
projection holds only in a Poincar\'e background, while in an AdS
background the chiral projection cannot be defined and we are
bounded to use 8-dimensional pseudo-Majorana spinors. In this
case the $R$-symmetry group reduces to the $SU(2)$ subgroup of
$SU(2)_L\otimes SU(2)_R$, the $R$-symmetry group of the chiral
spinors. For our purposes, it is convenient to use from the very
beginning 8-dimensional pseudo-Majorana spinors even in a
Poincar\'e framework, since we are going to discuss in a unique
setting both Poincar\'e and $AdS$ vacua.\\ The pseudo-Majorana
condition on the gravitino 1-forms is as follows:
\begin{equation}
(\psi_A)^{\dagger}\gamma^0=\overline{(\psi_A)}=\epsilon^{AB}\psi_{B}^{\
\ t}
\end{equation}
\noindent where we have chosen the charge conjugation matrix in
six dimensions as the identity matrix (an analogous definition
six dimensions as the identity matrix (an analogous definition
holds for the dilatino fields). We use eight dimensional
antisymmetric gamma matrices, with $(\gamma_7)^2=-1$. The indices
$A,B,\dots=1,2,$ of the spinor fields $\psi_A,\,\ \chi_A$
transform in the fundamental of the diagonal subgroup $SU(2)$ of
$SU(2)_L\otimes SU(2)_R$. To study the supersymmetric vacua let
us write down the Maurer-Cartan Equations (M.C.E.) dual to the
$F(4)$ Superalgebra (anti)commutators:
\begin{eqnarray}
\label{dVx}&&{\cal D}V^{a}-\frac{i}{2}\ \
\overline{\psi}_{A}\gamma_{a}\psi^A =0\\
&&{\mathcal R}^{ab}+4m^{2}\ \
V^{a}V^{b}+m\overline{\psi}_{A}\gamma_{ab}\psi^A=0\\
&&dA^{r}+\frac{1}{2}\,\ g\,\ \epsilon^{rst}A_{s}A_{t}-i\,\
\overline{\psi}_{A}\psi_{B}\,\
\sigma^{rAB}=0\\
\label{dPsix}&&D\psi_{A}-im\gamma_{a}\psi_{A}V^{a}=0
\end{eqnarray}

\noindent Here $V^a,\omega^{ab},\psi_A,A^r,(r= 1,2,3)$, are
superfield 1-forms dual to the $F(4)$ supergenerators  which at
$\theta =0$ have as $dx^{\mu}$ components
\begin{equation}
V^a_{\mu}= \delta^a_{\mu},\,\ \psi_{A\mu}=A^r_{\mu}=0, \,\
\omega^{ab}_{\mu}= pure\,\ gauge.
\end{equation}
Furthermore ${\cal R}^{ab}\equiv d\omega^{ab}-\omega^{ac}\land\,\
\omega_c^{\,\ b}$, ${\mathcal D}$ is the Lorentz covariant
derivative, $D$ is the $SO(1,5)\otimes SU(2)$ covariant
derivative, which on spinors acts as follows:
\begin{equation}D\psi_A\equiv
d\psi_A-\frac{1}{4}\gamma_{ab}\omega^{ab}\psi_A-\frac{i}{2}\sigma_{AB}^r
A_r\psi^B
\end{equation}
\noindent Note that $\sigma^{rAB}=\epsilon^{BC}\sigma^{rA}_{\ \
C}$, where $\sigma^{rA}_{\ \ B}$\ \ ($r=1,2,3$) denote the usual
Pauli matrices, are symmetric in $A,\,\ B$.\\ Let us point out
that the $F(4)$ superalgebra, despite the presence of two
different physical parameters, the $SU(2)$ gauge coupling constant
$g$ and the inverse $AdS$ radius $m$, really depends on just one
parameter since the closure under $d$-differentiation of eq.
(\ref{dPsix})  (equivalent to the implementation of Jacobi
identities on the generators), implies $g=3m$; to recover this
result one has to use the following Fierz identity involving
3-$\psi_A$'s 1-forms:
\begin{equation}
\label{fond}\frac{1}{4}\gamma_{ab}\psi_A\overline{\psi}_B\gamma^{ab}\psi_C\epsilon^{AC}-\frac{1}{2}
\gamma_{a}\psi_A\overline{\psi}_B\gamma^{a}\psi_C\epsilon^{AC}+3\psi_C\overline{\psi}_B\psi_A\epsilon^{BC}=0
\end{equation}\\The $F(4)$ superalgebra described by equations
(\ref{dVx}) -(\ref{dPsix}) fails to describe the physical vacuum
because of the absence of the superfields 2-form $B$ and 1-form
$A^0$ whose space-time restriction coincides with the physical
fields $B_{\mu\nu}$ and $A_{\mu}^0$ appearing in the supergravity
multiplet.
 The recipe to have all the fields in
a single algebra is well known and consists in considering the
Free Differential Algebra (F.D.A.)\cite{bible} obtained from the
$F(4)$ M.C.E.'s by adding two more equations for the 2-form $B$
and for the 1-form $A^0$ (the 0-form fields $\chi_A$ and $\sigma$
do not appear in the algebra since they are set equal to zero in
the vacuum). It turns out that to have a consistent F.D.A.
involving $B$ and $A^0$ one has to add to the $F(4)$ M.C.E.'s two
more equations involving $dA^0$ and $dB$; in this way one obtains
an extension of the M.C.E's to the following F.D.A:
\begin{eqnarray}\label{dV}&&{\mathcal D}V^{a}-\frac{i}{2}\ \
\overline{\psi}_{A}\gamma_{a}\psi^A=0 \\
\label{dO}&&{\mathcal R}^{ab}+4m^{2}\ \
V^{a}V^{b}+m\overline{\psi}_{A}\gamma_{ab}\psi^A=0\\
 \label{dAr}&&dA^{r}+\frac{1}{2}\,\ g\,\
\epsilon^{rst}A_{s}A_{t}-i\,\ \overline{\psi}_{A}\psi_{B}\,\
\sigma^{rAB}=0\\
\label{dA}&&dA^0-mB-i\,\
\overline{\psi}_{A}\gamma_{7}\psi^A=0\\
\label{dB}&&dB+2\,\
\overline{\psi}_{A}\gamma_{7}\gamma_{a}\psi^AV^{a}=0\\
\label{dPsi}&&D\psi_{A}-im\gamma_{a}\psi_{A}V^{a}=0\end{eqnarray}
\noindent  Equations (\ref{dA}) and (\ref{dB}) were obtained by
imposing that they satisfy the $d$-closure together with equations
(\ref{dV}). Actually the closure of (\ref{dB}) relies on the
4-$\psi_A$'s Fierz identity
\begin{equation}
\overline{\psi}_{A}\gamma_{7}\gamma_{a}\psi_{B}\epsilon^{AB}\overline{\psi}_{C}\gamma^{a}\psi_{D}\epsilon^{CD}=0
\end{equation}
The interesting feature of the F.D.A (\ref{dV})-(\ref{dPsi}) is
the appearance of the combination $dA^0-mB$ in (\ref{dA}). That
means that the dynamical theory obtained by gauging the F.D.A.
out of the vacuum will contain the fields $A^0_{\mu}$ and
$B_{\mu\nu}$ always in the single combination
$\partial_{[\mu}A^0_{\nu]}-mB_{\mu\nu}$. At the dynamical level
this implies, as noted by Romans \cite{rom}, an Higgs phenomenon
where the 2-form $B$ "eats" the 1-form $A^0$ and acquires a non
vanishing mass $m$.\\
It is easy to see that no F.D.A  exists if either $m=0$ , $g\neq
0$ or $m\neq 0$, $g= 0$, since the corresponding equations in the
F.D.A. do not close anymore under $d$- differentiation. In other
words the gauging of $SU(2)$, $g\neq 0$ must be necessarily
accompanied by the presence of the parameter $m$ which, as we have
 seen, makes the closure of the supersymmetric algebra consistent for $g=3m$.\\

In $D=6,\,\ N=4$ Supergravity, the only kind of matter is given
by vector multiplets, namely
\begin{equation}
(A_{\mu},\,\ \lambda_A,\,\ \phi^{\alpha})^I
\end{equation}
\noindent where $\alpha=0,1,2,3$ and the index $I$ labels an
arbitrary number $n$ of such multiplets. As it is well known the
$4n$ scalars parametrize the coset manifold $SO(4,n)/SO(4)\times
SO(n)$. Taking into account that the pure supergravity has a non
compact duality group $O(1,1)$ parametrized by $e^{\sigma}$, the
duality group of the matter coupled theory is
 \begin{equation}\label{coset}
  G/H=\frac{SO(4,n)}{SO(4)\times SO(n)}\times O(1,1)
\end{equation}
To perform the matter coupling we follow the geometrical
procedure of introducing the coset representative $L^{\Lambda}_{\
\ \Sigma}$ of the matter coset manifold, where
$\Lambda,\Sigma,\dots=0, \dots, 3+n$; decomposing the $O(4,n)$
indices with respect to $H=SO(4)\times O(n)$ we have:
\begin{equation}
L^{\Lambda}_{\ \ \Sigma}=(L^{\Lambda}_{\ \ \alpha},L^{\Lambda}_{\
\ I})
\end{equation}
\noindent where $\alpha=0,1,2,3$, $I=4,\dots ,3+n$. Furthermore,
since we are going to gauge the $SU(2)$ diagonal subgroup of
$O(4)$ as in pure Supergravity, we will also decompose
$L^{\Lambda}_{\ \ \alpha}$ as
\begin{equation}
L^{\Lambda}_{\ \ \alpha}=(L^{\Lambda}_{\ \ 0}, L^{\Lambda}_{\ \
r})
\end{equation}
The $4+n$ gravitational and matter vectors will now transform in
the fundamental of $SO(4,n)$ so that the superspace vector
curvatures will be now labeled by the index $\Lambda$:
$F^{\Lambda} \equiv (F^0,F^r,F^I)$. Furthermore the covariant
derivatives acting on the spinor fields will now contain also the
composite connections of the $SO(4,n)$ duality group. Let us
introduce the left-invariant 1-form of $SO(4,n)$ satisfying the
Maurer-Cartan equation
\begin{equation}
\Omega^{\Lambda}_{\ \ \Sigma}=(L^{\Lambda}_{\ \ \Pi})^{-1}
dL^{\Pi}_{\ \ \Sigma} \Rightarrow d\Omega^{\Lambda}_{\ \
\Sigma}+\Omega^{\Lambda}_{\ \ \Pi}\land\Omega^{\Pi}_{\ \ \Sigma}=0
\end{equation}

Our aim is to gauge a compact subgroup of $O(4,n)$. Since in any
case we may gauge only the diagonal subgroup $SU(2)\subset
O(4)\subset H$, the maximal gauging is given by
$SU(2)\otimes\mathcal{G}$ where $\mathcal{G}$ is a $n$-dimensional
subgroup of $O(n)$. According to a well known procedure, we
modify the definition of the left invariant 1-form $L^{-1}dL$ by
replacing the ordinary differential with the
$SU(2)\otimes\mathcal{G}$ covariant differential as follows:
\begin{equation}
\label{nabla}\nabla L^{\Lambda}_{\ \ \Sigma}=d L^{\Lambda}_{\ \
\Sigma}-f_{\Gamma\ \ \Pi}^{\,\ \Lambda} A^{\Gamma} L^{\Pi}_{\ \
\Sigma}
\end{equation}
\noindent where $f^{\Lambda}_{\ \ \Pi\Gamma}$ are the structure
constants of $SU(2)_d\otimes\mathcal{G}$. More explicitly,
denoting with $\epsilon^{rst}$ and $\mathcal{C}^{IJK}$ the
structure constants of the two factors $SU(2)$ and $\mathcal{G}$,
equation (\ref{nabla}) splits as follows:
\begin{eqnarray}
&&\nabla L^{0}_{\ \ \Sigma}=d L^{\Lambda}_{\ \ \Sigma}\\
&&\nabla L^{r}_{\ \ \Sigma}=d L^{r}_{\ \ \Sigma}-g\epsilon^{\,\
r}_{t\
\ s} A^{t} L^{s}_{\ \ \Sigma}\\
&&\nabla L^{I}_{\ \ \Sigma}=d L^{I}_{\ \ \Sigma}-g'{\mathcal
C}^{\,\ I}_{K\ \ J} A^{K} L^{J}_{\ \ \Sigma}
\end{eqnarray}
\noindent Setting $\widehat{\Omega}=L^{-1}\nabla L$, one easily
obtains the gauged Maurer-Cartan equations:
\begin{equation}\label{mc}d\widehat{\Omega}^{\Lambda}_{\ \
\Sigma}+\widehat{\Omega}^{\Lambda}_{\ \
\Pi}\land\widehat{\Omega}^{\Pi}_{\ \ \Sigma}=(L^{-1}{\mathcal
F}L)^{\Lambda}_{\ \ \Sigma}
\end{equation}
\noindent where $\cF\equiv\cF^{\Lambda}T_{\Lambda}$, $T_{\Lambda}$ being the generators of $SU(2)\otimes\cG$.\\
After gauging, by appropriate decomposition of the indices,we
find from (\ref{mc}):
\begin{eqnarray}
\label{11}&&R^r_{\,\ s}=-P^{r}_{\ \ I}\land P^I_{\ \ s}+(L^{-1}{\mathcal F}L)^{r}_{\ \ s}\\
\label{12}&&R^r_{\,\ 0}=-P^{r}_{\ \ I}\land P^I_{\ \ 0}+(L^{-1}{\mathcal F}L)^{r}_{\ \ 0}\\
\label{13}&&R^I_{\,\ J}=-P^I_{\ \ r}\land P^r_{\ \ J}-P^I_{\ \
0}\land P^0_{\ \ J}+(L^{-1}{\mathcal F}L)^{I}_{\ \ J}\\ \label{14}&&\nabla P^I_{\,\ r}=(L^{-1}{\mathcal F}L)^{I}_{\ \ r}\\
\label{15}&&\nabla P^I_{\,\ 0}=(L^{-1}{\mathcal F}L)^{I}_{\ \ 0}
\end{eqnarray}

\noindent Where we have set
\begin{displaymath}
P^I_{\alpha}=\left\{ \begin{array}{rr}P^I_{\,\ 0}\equiv
\Omega^{I}_{\ \ 0}\\ P^I_{\,\ r}\equiv \Omega^{I}_{\ \
r}\end{array}\right.
\end{displaymath}
\noindent Note that $P^I_0$, $P^I_r$ are the vielbeins of the
coset, while $(\Omega^{rs},\,\ \Omega^{r0})$, $(R^{rs},\,\
R^{ro})$ are respectively the connections and the curvatures of
$SO(4)$ decomposed with respect to the diagonal subgroup
$SU(2)\subset SO(4)$.\\
Starting from the F.D.A. discussed before, one can now define
suitable gauged curvatures in superspace, and apply the Bianchi
identities technique in order to retrieve the space-time
supersymmetry transformation laws. One finds:

{\setlength\arraycolsep{1pt}\begin{eqnarray} &\delta
V^{a}_{\mu}&=-i\overline{\psi}_{A\mu}\gamma^{a}\varepsilon^A\\
&\delta B_{\mu\nu}&=2 e^{-2\sigma}
\overline{\chi}_{A}\gamma_{7}\gamma_{\mu\nu}\varepsilon^A
-4e^{-2\sigma}\overline{\varepsilon}_A\gamma_7\gamma_{[\mu}\psi_{\nu]}^A\\
 &\delta A^{\Lambda}_{\mu}&=2 e^{\sigma}
 \overline{\varepsilon}^{A}\gamma_{7}\gamma_{\mu}\chi^BL^{\Lambda}_0\epsilon_{AB}+2e^{\sigma}\overline{\varepsilon}^{A}\gamma_{\mu}\chi^{B}L^{\Lambda r}\sigma_{rAB}-e^{\sigma}L_{\Lambda
I}\overline{\varepsilon}^{A}\gamma_{\mu}\lambda^{IB}\epsilon_{AB}+\nonumber\\&&+2ie^{\sigma}L^{\Lambda}_0\overline{\varepsilon}_A\gamma^7\psi_B\epsilon^{AB}+2ie^{\sigma}L^{\Lambda r}\sigma_{r}^{AB}\overline{\varepsilon}_A\psi_B\\
\label{qui}&\delta\psi_{A\mu}&={\mathcal
D}_{\mu}\varepsilon_A+\frac{1}{16}
e^{-\sigma}[T_{[AB]\nu\lambda}\gamma_{7}-T_{(AB)\nu\lambda}](\gamma_{\mu}^{\,\
\nu\lambda}-6\delta_{\mu}^{\nu}\gamma^{\lambda})
\varepsilon^{B}+\nonumber \\
 &&+\frac{i}{32}e^{2\sigma} H_{\nu\lambda\rho}
\gamma_{7}(\gamma_{\mu}^{\,\ \nu\lambda\rho}-3\delta_{\mu}^{\nu}
\gamma^{\lambda\rho})\varepsilon_{A}+\frac{1}{2}\varepsilon_{A}\overline{\chi}^{C}\psi_{C\mu}+\nonumber\\
&&+\frac{1}{2}\gamma_{7}\varepsilon_{A}\overline{\chi}^{C}\gamma^{7}\psi_{C\mu}-\gamma_{\nu}\varepsilon_{A}\overline{\chi}^{C}\gamma^{\nu}\psi_{C\mu}+\gamma_{7}\gamma_{\nu}\varepsilon_{A}\overline{\chi}^{C}\gamma^{7}\gamma^{\nu}\psi_{C\mu}+\nonumber\\
&&-\frac{1}{4}\gamma_{\nu\lambda}\varepsilon_{A}\overline{\chi}^{C}\gamma^{\nu\lambda}\psi_{C\mu}-\frac{1}{4}\gamma_{7}\gamma_{\nu\lambda}\varepsilon_{A}\overline{\chi}^{C}\gamma^{7}\gamma^{\nu\lambda}\psi_{C\mu}+{\bf S_{AB}}^{(g,g',m)}\gamma_{\mu}\varepsilon^B\\
\label{quo}&\delta\chi_{A}&=\frac{i}{2}
\gamma^{\mu}\partial_{\mu}\sigma \varepsilon_{A}\!+\!
\frac{i}{16}e^{-\sigma}[T_{[AB]\mu\nu}\gamma_{7}\!+\!T_{(AB)\mu\nu}]\gamma^{\mu\nu}\varepsilon^{B}\!+\!\frac{1}{32}e^{2\sigma}
H_{\mu\nu\lambda}\gamma_{7}\gamma^{\mu\nu\lambda}\varepsilon_{A}+\nonumber\\
&&+{\bf N_{AB}}^{(g,g',m)}\varepsilon^B\\
&\delta\sigma&=\overline{\chi}_{A}\varepsilon^A\\
\label{qua}&\delta\lambda^{IA}&=-iP^I_{ri}\sigma^{rAB}\partial_{\mu}\phi^{i}\gamma^{\mu}\varepsilon_{B}+iP^I_{0i}\epsilon^{AB}\partial_{\mu}\phi^{i}\gamma^{7}\gamma^{\mu}\varepsilon_{B}+\frac{i}{2}e^{-\sigma}T^{I}_{\mu\nu}\gamma^{\mu\nu}\varepsilon^{A}+\nonumber\\
&&+{\bf M_{AB}}^{I(g,g',m)}\varepsilon^B
\\
&P^{I}_{0i}\delta\phi^i&=\frac{1}{2}\overline{\lambda}^{I}_{A}\gamma_{7}\varepsilon^A\\
&P^{I}_{ri}\delta\phi^i&=\frac{1}{2}\overline{\lambda}^{I}_{A}\varepsilon_{B}\sigma_r^{ab}
\end{eqnarray}}
\noindent where we have introduced the "dressed" vector field
strengths
\begin{eqnarray}
&&T_{[AB]\mu\nu}\equiv\epsilon_{AB}L^{-1}_{0\Lambda
}F^{\Lambda}_{\mu\nu}\\
&&T_{(AB)\mu\nu}\equiv\sigma^r_{AB}L^{-1}_{r\Lambda}F^{\Lambda}_{\mu\nu}\\
&&T_{I\mu\nu}\equiv L^{-1}_{I\Lambda }F^{\Lambda}_{\mu\nu}
\end{eqnarray}
\noindent and we have omitted in the transformation laws of the
fermions the three-fermions terms of the form
$(\chi\chi\varepsilon)$, $(\lambda\lambda\varepsilon)$.\\ For our
purposes, the most interesting quantities are the fermionic
shifts denoted in boldface in the previous equations, since they
are the objects in terms of which one can retrieve the scalar
potential. Their explicit form is the following:

\begin{eqnarray}
\label{mS}&&
S_{AB}^{(g,g',m)}=\frac{i}{24}[Ae^{\sigma}+6me^{-3\sigma}(L^{-1})_{00})\epsilon_{AB}-\frac{i}{8}[B_te^{\sigma}-2me^{-3\sigma}(L^{-1})_{i0}]\gamma^7\sigma^t_{AB}\\
\label{mN}&&N_{AB}^{(g,g',m)}\!\!=\!\!\frac{1}{24}[Ae^{\sigma}-18me^{-3\sigma}(L^{-1})_{00})]\epsilon_{AB}+\frac{1}{8}[B_te^{\sigma}+6me^{-3\sigma}(L^{-1})_{i0}]\gamma^7\sigma^t_{AB}\\
\label{mM}&&M^{I(g,g',m)}_{AB}=(-C^I_t+2i\gamma^7D^I_t)e^{\sigma}\sigma^t_{AB}-2me^{-3\sigma}(L^{-1})^I_{\
\ 0}\gamma^7\epsilon_{AB}
\end{eqnarray}

\noindent where
\begin{equation}\label{AA}A=\epsilon^{rst}K_{rst};\ \ B^i=\epsilon^{ijk}K_{jk0};\ \ C_I^t=\epsilon^{trs}K_{rIs};\ \ D_{It}=K_{0It}\end{equation}
\noindent and the threefold completely antisymmetric tensors
$K's$ are the so called "boosted structure constants" given
explicitly by:
\begin{eqnarray}
&&K_{rst}=g\epsilon_{lmn}L^l_{\,\ r}(L^{-1})^{\,\ m}_sL^n_{\,\
t}+g'{\mathcal C}_{IJK}L^I_{\,\ r}(L^{-1})^{\,\ J}_sL^K_{\,\ t}\\
&&K_{rs0}=g\epsilon_{lmn}L^l_{\,\ r}(L^{-1})^{\,\ m}_sL^n_{\,\
0}+g'{\mathcal C}_{IJK}L^I_{\,\ r}(L^{-1})^{\,\ J}_sL^K_{\,\ 0}\\
&&K_{rIt}=g\epsilon_{lmn}L^l_{\,\ r}(L^{-1})^{\,\ m}_IL^n_{\,\
t}+g'{\mathcal C}_{LJK}L^L_{\,\ r}(L^{-1})^{\,\ J}_IL^K_{\,\ t}\\
&&K_{0It}=g\epsilon_{lmn}L^l_{\,\ 0}(L^{-1})^{\,\ m}_IL^n_{\,\
t}+g'{\mathcal C}_{LJK}L^L_{\,\ 0}(L^{-1})^{\,\ J}_IL^K_{\,\ t}
\end{eqnarray}

By performing the supersymmetry variation of the Lagrangian,
indeed one finds the following Ward identity \cite{fema}:
\begin{equation}
\label{ward}\delta^C_A{\mathcal
W}=-20S^{BA}S_{BC}-4N^{BA}N_{BC}+\frac{1}{4}M^{BA}_IM^I_{BC}
\end{equation}
\noindent One can verify that the r.h.s. of (\ref{ward}) is
indeed proportional to a Kronecker delta, by using the exlicit
form of the shifts. The resulting potential turns out to be:
{\setlength\arraycolsep{1pt}\begin{eqnarray}\label{pot}\mathcal{W}(\phi)=&&
5\,\ \{
[\frac{1}{12}(Ae^{\sigma}+6me^{-3\sigma}L_{00})]^2+[\frac{1}{4}(e^{\sigma}B_i-2me^{-3\sigma}L_{0i})]^2\}+\nonumber\\
\nonumber\\ &&- \{
[\frac{1}{12}(Ae^{\sigma}-18me^{-3\sigma}L_{00})]^2+[\frac{1}{4}(e^{\sigma}B_i+6me^{-3\sigma}L_{0i})]^2\}+\nonumber\\
&&-\frac{1}{4}\{C^I_{\,\ t}C_{It}+4D^I_{\,\ t}D_{It}\}\,\
e^{2\sigma}-m^2e^{-6\sigma}L_{0I}L^{0I}
\end{eqnarray}}

A further issue related to the scalar potential, which is an
important check of all our calculation, is the possibility of
computing the masses of the scalar fields by varying the
linearized kinetic terms  of the Lagrangian and the potential
${\mathcal W}$, after power expansion of $\mathcal{W}$ up to the
second order in the scalar fields $q^I_{\alpha}$. \noindent If we
use as mass unity the
\def\IP{\relax{\rm I\kern-.18em P}} inverse $AdS$ radius, which
in our conventions is $R^{-2}_{AdS}=4m^2$ we get:
\begin{equation}
\label{massa} m^2_{\sigma}=-6;\ \ m^2_{q^{I0}}=-4;\ \
m^2_{q^{Ir}}=-6
\end{equation}
\noindent These values should be compared with the results
obtained in reference \cite{fkpz} where the supergravity and
matter multiplets of the $AdS_6\,\ F(4)$ theory were constructed
in terms of the singleton fields of the 5-dimensional conformal
field theory, the singleton being given by hypermultiplets
transforming in the fundamental of $\mathcal{G}\equiv E_7$. It is
amusing to see that the values of the masses of the scalars
computed in terms of the conformal dimensions are exactly the
same as those given in equation (\ref{massa}).\\ This coincidence
can be considered as a non trivial check of the $AdS/CFT$
correspondence in six versus five dimensions.\\

\section{$F(4)\otimes\cG$ Superconformal Field Theory}
Here we describe the basics of the  $F(4)$ highest weight unitary
irreducible representations ``UIR's'' and exhibit two towers of
short representations which are relevant for a K-K analysis of
type IIA theory on  (warped) $AdS_6\otimes S^4$
\cite{oz},\cite{clp}.\\ We will not consider here the $\cG$
representation properties but we will only concentrate on the
supersymmetric structure.\\ Recalling that the even part of the
$F(4)$ superalgebra is $SO(2,5)\otimes SU(2)$, from a general
result on Harish-Chandra modules \cite{f}, \cite{h} of
$SO(2,2n+1)$ we know that there are only a spin 0 and a spin 1/2
singleton unitary irreducible representations \cite{flafro},
which, therefore, merge into a unique supersingleton
representation of the $F(4)$ superalgebra: the hypermultiplet
\cite{fkpz}.\\ To describe shortening is useful to use a harmonic
superfield language \cite{fanta1}.\\ The harmonic space is in
this case the 2-sphere $SU(2)/U(1)$, as in $N=2,\,\ d=4$ and
$N=1,\,\ d=6$. A highest weight UIR of $SO(2,5)$ is determined by
$E_0$ and a UIR of $SO(5)\simeq Usp(4)$, with Dynkin labels
$(a_1,\,\ a_2)$ \footnote {Note that the $Usp(4)$ Young labels
$h_1,h_2$ are related to $a_1,a_2$ by $a_1=2h_2; a_2=h_1-h_2$.}.
We will denote such representations by $\cD(E_0,\,\ a_1,\,\
a_2)$. The two singletons correspond to $E_0=3/2$, $a_1=a_2=0$
and $E_0=2$, $a_1=1$, $a_2=0$.\\ In the $AdS/CFT$ correspondence
\cite{malda, rass} $(E_0,\,\ a_1,\,\ a_2)$ become the conformal
dimension and the Dynkin labels of $SO(1,4)\simeq Usp(2,2)$.\\
The highest weight UIR of the $F(4)$ superalgebra will be denoted
by $\cD(E_0,\,\ a_1,\,\ a_2;\,\ I)$ where $I$ is the $SU(2)$
$R$-symmetry quantum number (integer or half integer).\\
The basic superfield is the supersingleton hypermultiplet
$W^A(x,\theta)$, which satisfies the constraint
\begin{equation}\label{hyper}
D_{\alpha}^{(A}W^{B)}(x,\theta)=0
\end{equation}
\noindent corresponding to the irrep. $\cD(E_0=\frac{3}{2},0,0;I=\frac{1}{2})$ \cite{fanta1}.\\
\noindent By using harmonic superspace, $(x,\,\ \theta_I,\,\
u^I_i)$, where $\theta_I=\theta_iu^i_I$, $u^i_I$ is the coset
representative of $SU(2)/U(1)$ and $I$ is the charge $U(1)$-label,
from  the covariant derivative algebra \begin{equation}
\{D^A_{\alpha},D^{jB}_{\beta}\}=i\epsilon^{AB}\partial_{\alpha\beta}
\end{equation}
 \noindent we have
\begin{equation} \{D^I_{\alpha},D^I_{\beta}\}=0 \ \ \ \
D^I_{\alpha}=D^i_{\alpha}u^I_i \end{equation} \noindent Therefore
from eq. (\ref{hyper}) it follows the $G$-analytic constraint:
\begin{equation} D_{\alpha}^{1}W^{1}=0\end{equation} \noindent which implies
\begin{equation}
W^1(x,\theta)=\varphi^1+\theta^{\alpha}_2\zeta_{\alpha}+d.t.
\end{equation}
 \noindent (d.t. means ``derivative terms'').\\
Note that $W^1$ also satisfies \begin{equation}
D^2_{\alpha}D^{2\alpha}W^1=0 \end{equation}
 \noindent because there is no such scalar component\footnote{This is rather similar to the treatment of the (1,0) hypermultiplet in $D=6$ \cite{fanta2}} in $W^1$.\\
$W^1$ is a Grassman  analytic  superfield, which is also harmonic
(that is ${\bf D}^1_2W^1=0$ where, using notations of reference
\cite{fesoca}, ${\bf D}^1_2$ is the step-up operator of the
$SU(2)$ algebra acting on  harmonic superspace).\\ Since $W^1$
satifies $D^1W^1=0$, any $p$-order polynomial \begin{equation}
\label{wp}I_p(W^1)=(W^1)^p\end{equation} \noindent will also have
the same property, so these operators form a ring under
multiplication \cite{fesoca}, they are the 1/2 BPS states of the
$F(4)$ superalgebra and represent massive vector
multiplets $(p>2)$, and massless bulk gauge fields for $p=2$.\\
The above multiplets correspond to the $D(E_0=3I,0,0;I=\frac{p}{2})$ h.w. U.I.R.'s of the $F(4)$ superalgebra.\\

The AdS squared mass for scalars is \begin{equation}
m_s^2=E_0(E_0-5) \end{equation}
 \noindent so there are three families of scalar states with
\begin{eqnarray*}
&&m^2_1=\frac{3}{4}p(3p-10)\hspace{20 mm}p\geq 2\\
&&m^2_2=\frac{1}{4}(3p+2)(3p-8)\hspace{10 mm}p\geq 2\\
&&m^2_3=\frac{1}{4}(3p+4)(3p-6)\hspace{10 mm}p\geq 4\\
\end{eqnarray*}
\noindent The only scalars states with $m^2<0$ are the scalar in
the massless vector multiplet $(p=2)$ with $m^2_1=-6$, $m^2_2=-4$
(no states with $m^2=0$ exist) and in the $p=3$ multiplet with
$m^2=-\frac{9}{4}$.\\ We now consider the second "short" tower
containing the graviton supermultiplet and its recurrences.\\ The
graviton multiplet is given by $W^1\overline{W}^1$. Note that
such superfield is not $G$-analytic, but it satisfies
\begin{equation}
D^1_{\alpha}D^{1\alpha}(W^1\overline{W}^1)=D^2_{\alpha}D^{2\alpha}(W^1\overline{W}^1)=0
\end{equation}
 \noindent this multiplet is the $F(4)$ supergravity multiplet.
 Its lowest component, corresponding to  the dilaton in $AdS_6$ supergravity multiplet,
 is a scalar with $E_0=3$ $(m^2=-6)$ and $I=0$.\\
 The tower is obtained as follows
\begin{equation} \label{magra}G_{q+2}(W)=W^1\overline{W}^1(W^1)^q \end{equation}
 \noindent where the massive graviton, described in eq..
 (\ref{magra}) has $E_0=5+\frac{3}{2}q$ and $I=\frac{q}{2}$.\\
 Note that the $G_{q+2}$ polynomial, although not $G$-analytic,
 satisfies the constraint
\begin{equation} D^1_{\alpha}D^{1\alpha}G_{q+2}(W)=0
\end{equation}
 \noindent so that it corresponds to a short representation with
 quantized dimensions and highest weight given by
 $D(E_0=3+3I,0,0;I=\frac{q}{2})$.\\
 We call these multiplets, following \cite{fanta2}, "intermediate
 short" because, although they have some missing states, they are
 not BPS in the sense of supersymmetry. In fact they do not form a
 ring under multiplication.\\
There are also long spin 2 multiplets containing $2^8$ state where $E_0$ is not quantized and satisfies the bound $E_0\geq 6$.\\
Finally let us make some comments on the role played by the flavour symmetry $\cG$.\\
It is clear that, since the supersingleton $W^1$ is in a
representation of $\cG$ (other than the gauge group of the
world-volume theory),
 the $I_p$ and $G_{q+2}$ polynomials will appear in the $p$-fold and
 $(q+2)$-fold tensor product representations of the $\cG$ group.
 This representation is in general reducible, however the 1/2 BPS
 states must have a first component totally symmetric in the $SU(2)$ indices  and, therefore, only
 certains $\cG$ representations survive.\\ Moreover in the $(W^1)^2$ multiplet,
 corresponding to the massless $\cG$- gauge vector multiplets in $AdS_6$, we must pick up the adjoint  representation
 $Adj\cG$ and in $W^1\overline{W}^1$, corresponding to the graviton multiplet, we must pick up  the $\cG$  singlet representation.\\
 However in principle there can be representations in the higher
 symmetric and antisymmetric products, and the conformal field theory should
  tell us which products remains, since the flavor symmetry
  depends on the specific dynamical model.\\
The states discussed in this paper are expected to appear
\cite{oz}, \cite{clp} in the
  K-K analysis of IIA massive supergravity on warped $AdS_6\otimes S^4$.

\end{document}